\documentclass[12pt]{article}
\usepackage{latexsym}
\usepackage{amsfonts}
\usepackage{amsmath}
\usepackage{amssymb}
\usepackage{graphicx}
\usepackage{latexsym}
\usepackage{amsfonts}
\usepackage{graphicx}
\usepackage{psfrag}

\input{tcilatex}
\begin{document}

\bigskip

\thispagestyle{empty}

\begin{center}
{\Large \textbf{Remark On Variance Bounds}}

\vskip0.2inR. Sharma\\[0pt]
Department of Mathematics \& Statistics\\[0pt]
Himachal Pradesh University\\[0pt]
Shimla -5,\\[0pt]
India - 171005\\[0pt]
email: rajesh\underline{~~}hpu\underline{~~}math@yahoo.co.in
\end{center}

\vskip1.5in \noindent \textbf{Abstract. }It is shown that the formula for
the variance of combined series yields surprisingly simple proofs of some
well known variance bounds. \textbf{\quad }



\bigskip

\bigskip

\bigskip

\bigskip

\bigskip

\bigskip

\bigskip

\bigskip

\bigskip

\bigskip

\bigskip

\bigskip

\bigskip

\bigskip

\bigskip

\bigskip

\bigskip

\bigskip

\bigskip

\bigskip

\bigskip

\section{Introduction}

\setcounter{equation}{0} It is well known that if we have two sets of data $%
X $ and $Y$ containing $n_{1}$ and $n_{2}$ observations with means $%
\overline{X}$ and $\overline{Y}$ , and variances $S_{n_{1}}^{2}$ and $%
S_{n_{2}}^{2},$ respectively, then the combined variance of $n_{1}+n_{2}$
observations is given by%
\begin{equation}
S_{n_{1}+n_{2}}^{2}=\frac{n_{1}}{n_{1}+n_{2}}S_{n_{1}}^{2}+\frac{n_{2}}{%
n_{1}+n_{2}}S_{n_{2}}^{2}+\frac{n_{1}n_{2}}{\left( n_{1}+n_{2}\right) ^{2}}%
\left( \overline{X}-\overline{Y}\right) ^{2}.  \tag{1.1}
\end{equation}

\noindent Let $X=\left\{ x_{j}\right\} $ be a sample of size one and let $Y$
be the sample of size $n-1$ drawn from the population $\left\{
x_{1},x_{2},...,x_{n}\right\} $ such that $X\cap Y=\phi .$ Then $%
n_{1}=1,n_{2}=n-1,\overline{X}=x_{j},\overline{Y}=\frac{1}{n}%
\sum\limits_{i\neq j}x_{j},S_{1}^{2}=0,$ and it follows from (1.1)
that\noindent 
\begin{equation}
S_{n}^{2}=\frac{n-1}{n}S_{n-1}^{2}+\frac{1}{n-1}\left( x_{j}-\overline{x}%
\right) ^{2},  \tag{1.2}
\end{equation}%
where $\overline{x}=\frac{1}{n}\sum\limits_{i=1}^{n}x_{i}$ and $S_{n}^{2}=%
\frac{1}{n}\sum\limits_{i=1}^{n}\left( x_{i}-\overline{x}\right) ^{2}$are
respectively the mean and variance of the data $\left\{
x_{1},x_{2},...,x_{n}\right\} .$

\noindent Each summand in (1.2) is non-negative, so%
\begin{equation}
S_{n}^{2}\geq \frac{1}{n-1}\left( x_{j}-\overline{x}\right) ^{2},  \tag{1.3}
\end{equation}%
for all $j=1,2,...,n.$

\noindent The inequality (1.3) is known as Samuelson's inequality (1968) in
statistical literature. The inequality (1.3) was also established in
mathematical literature by Laguerre (1880) in some different context and
notations. Several alternative proofs of this inequality were given in
literature, see Arnold and Balakrishna (1989), and Rassias and Srivastava
(1999).

\noindent It may be noted here that the identity (1.2) also implies that 
\begin{equation*}
S_{n}^{2}\geq \frac{n-1}{n}S_{n-1}^{2}.
\end{equation*}%
Thus, if $S_{m}^{2}$ is the variance of a sample of size $m$ drawn from a
population of size $n,$ then%
\begin{equation*}
S_{n}^{2}\geq \frac{m}{n}S_{m}^{2}.
\end{equation*}%
Let $X=\left\{ x_{j},x_{k}\right\} $ be a sample of size $2$ and let $Y$ be
the sample of size $n-2$ drawn from the population $\left\{
x_{1},x_{2},...,x_{n}\right\} $ such that $X\cap Y=\phi .$ Then $n_{1}=2,$ $%
n_{2}=n-2,$ $\overline{X}=\frac{x_{j}+x_{k}}{2},$ $\overline{Y}=\frac{1}{n-2}%
\sum\limits_{i\neq j,k}x_{i},$ $S_{2}^{2}=\frac{\left( x_{j}-x_{k}\right)
^{2}}{4},$ and it follows from (1.1) that%
\begin{equation}
S_{n}^{2}=\frac{n-2}{2}S_{n-2}^{2}+\frac{1}{2n}\left( x_{j}-x_{k}\right)
^{2}+\frac{2}{n-2}\left( \overline{x}-\frac{x_{j}+x_{k}}{2}\right) ^{2}, 
\tag{1.4}
\end{equation}%
for all $j=1,2,...,n$ and $n\geq 3.$

\noindent Each summand in (1.4) is non-negative, so%
\begin{equation}
S_{n}^{2}\geq \frac{1}{2n}\left( x_{j}-x_{k}\right) ^{2},  \tag{1.5}
\end{equation}%
for all $j=1,2,...,n.$ From (1.5), for $m\leq x_{i}\leq M,i=1,2,...,n,$ we
have%
\begin{equation}
S_{n}^{2}\geq \frac{1}{2n}\left( M-m\right) ^{2}.  \tag{1.6}
\end{equation}%
The inequality (1.6) is due to Nagy (1918). See also Nair (1948) and
Thompson (1935).

\noindent Likewise, from (1.4), we have for $n\geq 3,$%
\begin{equation}
S_{n}^{2}\geq \frac{1}{2n}\left( M-m\right) ^{2}+\frac{2}{n-2}\left( 
\overline{x}-\frac{m+M}{2}\right) ^{2}.  \tag{1.7}
\end{equation}%
The inequality (1.7) provides a refinement of (1.6), see Sharma et al.
(2008).

\noindent Mallows and Richter (1969) proves an extension of the Samuelson
inequality (1.3). This says that if $\gamma _{r}$ is the mean of any subset
of $r$ numbers chosen from the set $\left\{ x_{1},x_{2},...,x_{n}\right\} ,$
then%
\begin{equation}
S_{n}^{2}\geq \frac{r}{n-r}\left( \gamma _{r}-\overline{x}\right) ^{2}, 
\tag{1.8}
\end{equation}%
for $r=1,2,...,n-1.$

\noindent From (1.1), we have 
\begin{equation}
S_{n_{1}+n_{2}}^{2}\geq \frac{n_{1}n_{2}}{\left( n_{1}+n_{2}\right) ^{2}}%
\left( \overline{X}-\overline{Y}\right) ^{2}.  \tag{1.9}
\end{equation}%
Let $X$ be a sample of size $r$ and let $Y$ be the sample of size $n-r$
drawn from the population $\left\{ x_{1},x_{2},...,x_{n}\right\} $ such that 
$X\cap Y=\phi .$ Then $n_{1}=r,$ $n_{2}=n-r,$ $\overline{X}-\overline{Y}=%
\frac{n}{n-r}\left( \gamma _{r}-\overline{x}\right) ,$and so (1.8) follows
from (1.9).

\noindent Likewise, we can deduce Boyd-Hawkins inequalities (1971) from
(1.9). This says that if $x_{1}\leq x_{2}\leq ...\leq x_{n},$ then%
\begin{equation*}
\overline{x}-\sqrt{\frac{n-k}{k}}S_{n}\leq x_{k}\leq \overline{x}+\sqrt{%
\frac{k-1}{n-k+1}}S_{n}
\end{equation*}%
for $k=2.3,...,n-1.$

\noindent The variance bounds have various extensions and applications in
statistics, polynomials and matrix theory. We see that formula (1.1)
provides further insight, and is very useful in the study of these
inequalities. In this way we can study various further refinements,
generalisations and extensions of the variance bounds.

\end{document}